\documentstyle[preprint,eqsecnum,aps]{revtex}

\begin{document}

\title {Quantization of a generally covariant gauge system
with two super Hamiltonian constraints}

\author{Rafael Ferraro{\footnote{Electronic address:
ferraro@iafe.uba.ar}} and
Daniel M. Sforza{\footnote{Electronic address:
sforza@iafe.uba.ar}}}

\address{
{\it Instituto de Astronom\'\i a y F\'\i sica del
Espacio,\\ Casilla de Correo
67, Sucursal 28, 1428 Buenos Aires, Argentina\\ and
Departamento de F\'\i sica,
Facultad de Ciencias Exactas y Naturales,\\ Universidad de
Buenos Aires, Ciudad
Universitaria, Pab. I, 1428 Buenos Aires, Argentina\\}}

\maketitle

\begin{abstract}

The Becci-Rouet-Stora-Tyutin (BRST) operator quantization of a
finite-dimensional gauge system featuring two quadratic super Hamiltonian and
$m$ linear supermomentum constraints is studied as a model for quantizing
generally covariant gauge theories. The proposed model ``completely'' mimics
the constraint algebra of general relativity. The Dirac constraint operators
are identified by realizing the BRST generator of the system as a Hermitian
nilpotent operator, and a physical inner product is introduced to complete a
consistent quantization procedure.

\end{abstract}
\vskip  1cm


\newpage

\section{Introduction}

Generally covariant theories such as Einstein's theory of gravitation have the
peculiar property of featuring a Hamiltonian that is constrained to vanish.
This constraint is associated with the invariance of the action under
reparametrizations. In the case of general relativity, the
Arnowitt-Deser-Misner (ADM) Hamiltonian is the sum of four constraint
functions, three of them are the supermomenta ${\mathcal H}_a$ (linear and
homogeneous functions of the field momenta) and the other one is the
super-Hamiltonian ${\mathcal H}$ (a quadratic function of the field momenta).
The supermomenta generate the change of the field under diffeomorphisms on the
spatial hypersurfaces $\Sigma$ of the foliated space-time manifold, and express
the invariance of the action under these transformations. The super-Hamiltonian
generates the evolution of the field under displacements that are normal to the
hypersurface $\Sigma$; this evolution is nothing but a physically irrelevant
reparametrization of the dynamical trajectory of the system. Actually there are
four constraint functions in each point of $\Sigma$. So the algebra of
constraints is rather complicated, because one should evaluate the Poisson
brackets between constraints at different points. In particular, the Poisson
brackets between two super-Hamiltonian take part in the algebra of constraints.
This algebra was calculated by Dirac\cite{dirac}:
\begin{eqnarray}\label{pb}
\lefteqn{\{{\mathcal H}(x),{\mathcal H}(x')\}={\mathcal
H}^a(x)\delta_{,a}(x,x')+ {\mathcal
H}^a(x')\delta_{,a}(x,x'),} \\
&&\{{\mathcal H}_a(x),{\mathcal H}(x')\}={\mathcal
H}(x)\delta_{,a}(x,x'),\label{pb2}\\ &&\{{\mathcal
H}_a(x),{\mathcal
H}_b(x')\}={\mathcal H}_b(x)\delta_{,a}(x,x')+ {\mathcal
H}_a(x')\delta_{,b}(x,x'),\label{pb3}
\end{eqnarray}
and his first class character guarantees that the dynamical
trajectories are
consistent with the constraints
\begin{equation}
{\mathcal H}=0,\ \ \ \ \ {\mathcal H}_a=0.
\end{equation}
According to Dirac, the quantization of a constrained system requires a factor
ordering able to preserve the algebra of constraints at the level of operators
(absence of anomalies). In the case of general relativity this issue remains
unsolved \cite{v1,v2}. In order to avoid the regularization of operators, it is
a common practice to study this question in finite-dimensional systems
featuring constraints that resemble the algebra of general relativity. Although
these types of systems have been widely studied (see, for example, Refs.
\cite{hk90,fs97,fs99,fs00}), it has been pointed out by Montesimos et
al.\cite{mrt99} that an important feature of the algebra (\ref{pb})-(\ref{pb3})
has not been sufficiently examined: usually people does not deal with Eq.
(\ref{pb}) but only with (\ref{pb2}),(\ref{pb3}), i.e. frequently only one
super-Hamiltonian is included (however, models with several {\it commuting}
Hamiltonian constraints were considered in Refs. \cite{lo86,lo89,lu97}).
Actually, Montesinos et al.\cite{mrt99} solved a simple model with two
euclidean flat super-Hamiltonians plus a  single supermomenta satisfying only
Eqs. (\ref{pb})-(\ref{pb2}) with constant structure coefficients; thanks to the
simple structure of the constraints, the ordering problem is trivially solved
in this case. Our aim is to quantize a more complex system in order to deal
with a non trival ordering of constraint operators that mimic the complete
algebra (\ref{pb})-(\ref{pb3}).

It is well known that, besides the Dirac method, the Becchi-Rouet-Stora-Tyutin
(BRST) formalism is a powerful tool to quantize a first class constrained
system \cite{ht} (see also Ref. \cite{bfv} for an account of the
Batalin-Fradkin-Vilkovisky formalism). In recent works \cite{fs97,fs99,fs00},
we have taken advantage of its strength for obtaining the consistently ordered
constraint operators belonging to generally covariant systems including only
one super-Hamiltonian constraint. Here, we will show that the tools there
developed are also useful in the treatment of a nontrivial system featuring
{\it two} super-Hamiltonian constraints an $m$ supermomenta constraints.

We will start by defining the model. Then, the system will be quantized within
the framework of the BRST formalism, where the nilpotency of the BRST generator
must be proven in order to guarantee an anomaly free quantization. The Dirac
constraint operators will be identified from the nilpotent BRST generator. In
spite of this quantization will be performed for a system featuring an algebra
with constant structure functions, the results will be extended to more general
algebras by means of a unitary transformation. Finally, a physical inner
product will be introduced to complete a consistent quantization procedure.

\section{The model}

Let us consider a system described by $4n$ canonical
coordinates $(q^i,p_i)$,
subjected to {\it two} super-Hamiltonian constraints

\begin{equation}\label{h1}
{\cal H}_1={1\over 2}\
g^{i_1j_1}(q^{k_1})p_{i_1}p_{j_1}+{\upsilon}_1(q^{k_2}),
\end{equation}
\begin{equation}\label{h2}
{\cal H}_2={1\over 2}\
g^{i_2j_2}(q^{k_2})p_{i_2}p_{j_2}+{\upsilon}_2(q^{k_1}),
\end{equation}
where $i_1=1,...,n$ and $i_2=n+1,...,2n$.
\medskip

The metrics $g^{i_1j_1}$ and $g^{i_2j_2}$ are indefinite and non degenerated,
and depend on the $q^{i_1}$'s and the $q^{i_2}$'s respectively. On the
contrary, the potentials exhibit an opposite functional dependence:
${\upsilon}_1={\upsilon}_1(q^{i_2})$ and ${\upsilon}_2={\upsilon}_2(q^{i_1})$.

In addition, the system is also subjected to $m$ linearly independent
supermomentum constraints
\begin{equation}\label{g}
{\cal H}_a=\xi_a^i\ p_i,  ~~~~~~a=3,...,m+2; ~~i=(i_1,i_2),
\end{equation}
where
\begin{equation}\label{proplin}
\xi_a^{i_1}=\xi_a^{i_1}(q^{k_1}),~~
\xi_a^{i_2}=\xi_a^{i_2}(q^{k_2}).
\end{equation}
The special way the geometrical objects in the constraint functions depend on
the coordinates, has been chosen for obtaining a constraint algebra which
mimics Eqs. (\ref{pb})-(\ref{pb3}):

\begin{equation}\label{chh}
\{{\cal H}_1 ,{\cal H}_2 \}=c_{12}^a{\cal H}_a,
\end{equation}
\begin{equation}\label{ch1g}
\{{\cal H}_1,{\cal H}_a\}=c_{1a}^1 {\cal H}_1,
\end{equation}
\begin{equation}\label{ch2g}
\{{\cal H}_2,{\cal H}_a\}=c_{2a}^2 {\cal H}_2,
\end{equation}
\begin{equation}\label{cgg}
\{ {\cal H}_a,{\cal H}_b \}=C_{ab}^c {\cal H}_c.
\end{equation}

It should be noticed that this system describes two {\it interacting}
particles. In fact, it is not possible to decouple the system in two subsystems
described by $(q^{i_1}, p_{i_1})$ and $(q^{i_2}, p_{i_2})$. As long as we know,
models of this class in riemannian manifolds have not been studied yet in the
literature.

We will start by quantizing an algebra with constant structure functions.
Later, by taking into account the scaling properties of the super-Hamiltonians,
we will extend the results to some algebras with non constant structure
functions.

In order that the constraints (\ref{h1}), (\ref{h2}), and (\ref{g}) effectively
satisfy the algebra (\ref{chh})-(\ref{cgg}), metrics, vectors and potentials
must fulfill certain relations. By substituting the constraints in Eq.
(\ref{chh}) one obtains
\begin{equation}\label{chhcond}
 g^{i_2k_2} {\upsilon}_{1,k_2}p_{i_2} -g^{i_1k_1}
{\upsilon}_{2,k_1}p_{i_1}=c_{12}^a(
\xi_a^{i_1}p_{i_1}+\xi_a^{i_2}p_{i_2}),
\end{equation}
then,\footnote{It must be noticed that Eqs. (2.10) and
(2.11) do not force
$c_{12}^a$ to be a constant; they only imply that
$c_{12}^a=~^{(1)}c_{12}^a(q^{i_1})-~^{(2)}c_{12}^a(q^{i_2})$,
where
$~^{(1)}c_{12}^a\xi_a^{i_1}=0$ and
$~^{(2)}c_{12}^a\xi_a^{i_2}=0$.}
\begin{equation}\label{vcg61}
c_{12}^a\xi_a^{i_2}=g^{i_2k_2} {\upsilon}_{1,k_2}
\end{equation}
and
\begin{equation}\label{vcg62}
c_{12}^a\xi_a^{i_1}=-g^{i_1k_1} {\upsilon}_{2,k_1}.
\end{equation}

The substitution of the constraints in Eq. (\ref{ch1g}) yields
\begin{equation}\label{ch1gcond}
{1\over 2} (g^{i_1j_1}_{,k_1}\xi_a^{k_1}
-2g^{i_1k_1}\xi_{a,k_1}^{j_1})p_{i_1}p_{j_1}+
{\upsilon}_{1,k_2}\xi_a^{k_2}=c_{1a}^1(g^{i_1j_1}p_{i_1}p_{j_1}+
{\upsilon}_1)
\end{equation}
then
\begin{equation}\label{cond611}
g^{i_1j_1}_{,k_1}\xi_a^{k_1}-2g^{i_1k_1}\xi_{a,k_1}^{j_1}=c_{1a}^1g^{i_1j_1}
\end{equation}
and
\begin{equation}\label{cond612}
\xi_a^{k_2}{\upsilon}_{1,k_2}=c_{1a}^1 {\upsilon}_1.
\end{equation}
In geometrical language, the relations (\ref{cond611}) and
(\ref{cond612}) read
\begin{eqnarray}\label{lie1}
&&{\cal L}_{\vec\xi_a}\bar{\bar{g_1}}=c_{1a}^1\bar{\bar{g_1}}, \\ &&{\cal
L}_{\vec\xi_a} {\upsilon}_1=c_{1a}^1 {\upsilon}_1.
\end{eqnarray}

Analogously, the substitution of the constraints in Eq. (\ref{ch2g}) leads to
\begin{equation}\label{ch2gcond}
{1\over 2}(g^{i_2j_2}_{,k_2}\xi_a^{k_2}
-2g^{i_2k_2}\xi_{a,k_2}^{j_2})p_{i_2}p_{j_2}+
{\upsilon}_{2,k_1}\xi_a^{k_1}=c_{2a}^2(g^{i_2j_2}p_{i_2}p_{j_2}+
{\upsilon}_2)
\end{equation}
which means
\begin{equation}\label{cond621}
g^{i_2j_2}_{,k_2}\xi_a^{k_2}-2g^{i_2k_2}\xi_{a,k_2}^{j_2}=c_{2a}^2g^{i_2j_2}
\end{equation}
and
\begin{equation}\label{cond622}
\xi_a^{k_1}{\upsilon}_{2,k_1}=c_{2a}^2 {\upsilon}_2
\end{equation}
or, in geometrical language,
\begin{eqnarray}\label{lie2}
&&{\cal L}_{\vec\xi_a}\bar{\bar{g_2}}=c_{2a}^2\bar{\bar{g_2}}, \\ &&{\cal
L}_{\vec\xi_a} {\upsilon}_2=c_{2a}^2 {\upsilon}_2.
\end{eqnarray}
Thus, the fulfillment of the algebra means that the supermomenta are conformal
Killing vectors of the super-Hamiltonians.

Finally, by substituting the supermomenta in Eq. (\ref{cgg}) one obtains that
\footnote{Eq. (2.22) does not imply that the $C_{ab}^c$'s are constant, it only
imposes the decomposition:
$C_{ab}^c=~^{(1)}C_{ab}^c(q^{i_1})-~^{(2)}C_{ab}^c(q^{i_2})$ where
$~^{(1)}C_{ab}^c\xi_c^{i_1}=0$ and $~^{(2)}C_{ab}^c\xi_c^{i_2}=0$.}
\begin{equation}\label{cggcond}
(\xi_a^{i_1}\xi_{b,i_1}^{j_1}-\xi_b^{i_1}\xi_{a,i_1}^{j_1})\
p_{j_1}+
(\xi_a^{i_2}\xi_{b,i_2}^{j_2}-\xi_b^{i_2}\xi_{a,i_2}^{j_2})\
p_{j_2}= C_{ab}^c (
\xi_c^{i_1}\ p_{i_1}+\xi_c^{i_2}\ p_{i_2}).
\end{equation}

\section{The BRST formalism: classical and quantum
generators}

Our aim is to find the constraint operators that satisfy the algebra at the
quantum level. The appropriate factor ordering can be found within the
framework of the BRST formalism. For this, the original phase space is extended
by including a canonically conjugate pair of fermionic ghosts $(\eta^a,{\cal
P}_a)$ for each constraint function. The central object is the BRST generator,
a fermionic function $\Omega=\Omega(q^i,p_j,\eta^a,{\cal P}_b)$ that captures
all of identities satisfied by the set of first class constraints in a unique
identity
\begin{equation}\{\Omega,\Omega\}=0.\label{nil}\end{equation}

The existence of $\Omega$ is guaranteed at the classical level, and $\Omega$ is
unique up to canonical transformations in the extended phase space.  It can be
built by means of a recursive method \cite{ht}.  In the present case the result
is (a closed algebras has ``rank" equal to 1)
\begin{equation}
\Omega=\eta^1 {\cal H}_1 +\eta^2 {\cal H}_2 + \eta^a  {\cal
H}_a + \eta^1
\eta^2 c_{12}^a {\cal P}_a + \eta^1 \eta^a c_{1a}^1 {\cal
P}_1 +  \eta^2 \eta^a
c_{2a}^2 {\cal P}_2 + {1\over  2} \eta^a  \eta^b C_{ab}^c
{\cal
P}_c.\label{omegaclm}
\end{equation}

In order to quantize the extended system, the classical BRST
generator must be realized as a Hermitian operator. The theory is
free from BRST anomalies, if a Hermitian realization of $\Omega$
can be found such that the classical property (\ref{nil}) becomes
\begin{equation}
[{\hat\Omega},{\hat\Omega}]=2{\hat\Omega}^2=0
,\label{nilq}\end{equation} i.e.,
${\hat{\Omega}}$ must be nilpotent.  The BRST physical
quantum states belong to
the set of equivalence classes of BRST-closed states
(${\hat{\Omega}}\psi=0$)
moduli BRST-exact ones ($\psi={\hat{\Omega}}\chi$) (quantum
BRST cohomology).

In order to get the Hermitian and nilpotent operator $\hat\Omega$ is helpful to
write $\Omega$ with the canonical and ghost momenta on an equal footing. So,
let us adopt the notation
\begin{equation}
\eta^{C_s}=(q^i,\eta^{\beta}),
~~~~~~~~~~~~{\cal P}_{C_s}=(p_i,{\cal P}_{\beta}),
\end{equation}
where $s=-1,0$, and ${\beta}=(A;a)=(1,2;3,...,m+2)$. The index $s$
distinguishes original variables from the added ghost variables.

Also, $\Omega$ is a sum of a term quadratic in the momenta
\begin{equation}
{\Omega}^{quad}=\eta^A {\cal H}_A \equiv{1\over 2}
\sum_{r,s=-1}^0 \Omega^{A_rB_s}{\cal P}_{A_r} {\cal
P}_{B_s}+\eta^A\upsilon_A,
\end{equation}
plus another term linear in the momenta
\begin{equation}\Omega^{linear}= \eta^a  {\cal H}_a +
\eta^1 \eta^2 c_{12}^a
{\cal P}_a + \eta^1 \eta^a c_{1a}^1 {\cal P}_1 +  \eta^2
\eta^a c_{2a}^2 {\cal
P}_2 + {1\over  2} \eta^a  \eta^b C_{ab}^c {\cal
P}_c\equiv\sum_{s=-1}^0
\Omega^{c_s}{\cal P}_{c_s}. \label{omlin6}\end{equation}

In general, it is a difficult step to get the Hermitian nilpotent BRST operator
from its classical counterpart because a general method is not available (when
not unrealizable: at the quantum level its mere existence is not guaranteed).
Nevertheless, let us begin by proposing for the linear term the operator
\cite{fs97,fs99}
\begin{equation}
\hat{\Omega}^{linear}=\sum_{s=-1}^0f^{1\over
2}\hat\Omega^{c_s} \hat{\cal
P}_{c_s}f^{-{1\over 2}}\label{ordlin6}\end{equation}
where $f=f(q^i)$ depends on all of original coordinates.

$\hat{\Omega}^{linear}$ will be Hermitian if $f$ satisfies
\begin{equation}C^{\beta}_{a\beta}=f^{-
1}(f\xi^i_a)_{,i},\label{div6}\end{equation}
where $C^{\beta}_{a\beta}=C^b_{ab}+c^1_{a1}+c^2_{a2}$ (please remember that
 $\beta$ runs over all constraint labels). In
 Eq. (\ref{div6}), $f$ behaves as a volume in the gauge orbit of the
supermomenta (see Ref. \cite{fs97}).

The quadratic term can be caught in the Hermitian ordering:
\begin{equation}\hat{\Omega}^{quad}={1\over 2}
\sum_{r,s=-1}^0 f^{-{1\over  2}}\hat{\cal  P}_{A_r} f
\Omega^{A_rB_s} \hat{\cal
P}_{B_s}f^{-{1\over 2}}+\hat \eta^A\upsilon_A.
\label{ordcuad6}\end{equation}

Finally, the proposed Hermitian BRST operator can be rearranged in the $\hat
\eta-\hat {\cal P}$ order by repeatedly using the ghost (anti)commutation
relations. After this procedure is completed, the classical structure of Eq.
(\ref{omegaclm}) will be reproduced at the quantum level, although it will be
free of anomalies only if $\hat \Omega$ is nilpotent\cite{ht}
\begin{equation}
\hat\Omega={\hat\eta}^1 \hat{\cal H}_1+ {\hat\eta}^2
\hat{\cal H}_2+ \hat \eta^a
\hat {\cal H}_a+\hat \eta^1 \hat\eta^2 c_{12}^a \hat {\cal
P}_a +\hat\eta^1
\hat\eta^a c_{1a}^1 \hat {\cal P}_1+ \hat\eta^2  \hat\eta^a
c_{2a}^2 \hat{\cal
P}_2 + {1\over 2} \hat \eta^a \hat \eta^b C_{ab}^c \hat
{\cal P}_c.
\label{omeord'm}\end{equation}

In this case, the result is

\begin{equation}\label{h1c}
\hat{\cal H}_1=\frac{1}{2}  f^{-{1\over 2}} \hat p_{i_1}
g^{i_1j_1} f \hat
p_{j_1} f^{-{1\over 2}}+ {\upsilon}_1,
\end{equation}
\begin{equation}\label{h2c}
\hat{\cal H}_2=\frac{1}{2}  f^{-{1\over 2}} \hat p_{i_2} g^{i_2j_2} f \hat
p_{j_2} f^{-{1\over 2}}+ {\upsilon}_2,
\end{equation}

\begin{equation}\label{gc}
\hat {\cal H}_a=f^{1\over 2} \xi^i_a \hat p_i f^{-{1\over
2}}.
\end{equation}

We still need to demand the nilpotency of the proposed $\hat \Omega$. We expect
additional conditions over $f(q^i)$ since Eq. (\ref{div6}) does not completely
fix it.

{\it Proof of} $\hat\Omega^2=0$. The proof is done by explicit calculation.
Here we give an abridged demonstration; the full version can be found in Ref.
\cite{tesis}. After the removing of the terms that cancel out due to the Jacobi
identities such as

\begin{eqnarray}\label{idj}
\lefteqn{\{\{{\cal H}_1,{\cal H}_2\},{\cal H}_a\}+\{\{{\cal
H}_2,{\cal
H}_a\},{\cal
H}_1\}+\{\{{\cal H}_a,{\cal H}_1\},{\cal H}_2\}=0}\nonumber
\\ &&\Longrightarrow
c_{12}^b
C_{ba}^c- c_{12}^c (c_{1a}^1+c_{2a}^2)=0,
\end{eqnarray}
or that are identically null [$(\eta^a)^2\equiv 0$, etc.], one finally gets
\begin{eqnarray}\label{nilpfin}
\lefteqn{\!\!\!\!\!\!\!\![\hat \Omega,\hat \Omega]= 2 {\hat\eta}^1
{\hat\eta}^2([\hat{\cal H}_1, \hat{\cal H}_2]-i c_{12}^b \hat {\cal H}_b)}
\nonumber \\ && +2{\hat\eta}^1 {\hat\eta}^a([ \hat{\cal H}_1, \hat{\cal H}_a]-i
c_{1a}^1\hat{\cal H}_1) \nonumber \\ && + 2{\hat\eta}^2 {\hat\eta}^a([
\hat{\cal H}_2, \hat{\cal H}_a] -i c_{2a}^2\hat{\cal H}_2).
\end{eqnarray}

Therefore the nilpotency is apparent whenever the constraint operators
$\hat{\cal H}_1,~\hat{\cal H}_2$, and $\hat{\cal H}_a$ realize the first class
constraint algebra at the quantum level:
\begin{equation}\label{q12}
[\hat{\cal H}_1,\hat{\cal H}_2]=~ i c_{12}^a \hat {\cal
H}_a,
\end{equation}
\begin{equation}\label{q1a}
\ \ \ \ \ \ \ \ \ [\hat{\cal H}_A,\hat{\cal H}_a]=~ i
c_{Aa}^{(A)} \hat {\cal
H}_{(A)}
\end{equation}
(there is no sum over label $A$). Then, let us calculate
explicitly Eq.
(\ref{q12}):
\begin{eqnarray}\label{ccord12}
\lefteqn{\!\!\!\!\!\!\!\!\!\!\!\!\! [\hat{\cal
H}_1,\hat{\cal
H}_2]=[\frac{1}{2} f^{-{1\over 2}} \hat p_{i_1} g^{i_1j_1}
f \hat p_{j_1}
f^{-{1\over 2}}+ {\upsilon}_1,\frac{1}{2} f^{-{1\over 2}}
\hat p_{i_2}
g^{i_2j_2} f \hat p_{j_2} f^{-{1\over 2}}+
{\upsilon}_2]}\nonumber \\
&&=f^{1\over 2}  [\frac{1}{2} f^{-1} \hat p_{i_1}
g^{i_1j_1} f \hat p_{j_1}+
{\upsilon}_1,\frac{1}{2} f^{-1} \hat p_{i_2} g^{i_2j_2} f
\hat p_{j_2}+
{\upsilon}_2] f^{-{1\over 2}}\nonumber
\\ &&=f^{1\over 2}  [\frac{1}{2}g^{i_1j_1} \hat p_{i_1}
\hat p_{j_1} -{i\over 2} f^{-1} (g^{i_1j_1} f)_{,i_1} \hat p_{j_1} +
{\upsilon}_1,\nonumber \\ && \ \ \ \ \ \ \ \ \ \ \ \ \ \ \ \ \ \ \ \ \ \ \ \ \
\ \ \ \ \ ,\frac{1}{2} g^{i_2j_2} \hat p_{i_2} \hat p_{j_2}-{i\over 2} f^{-1}
(g^{i_2j_2} f)_{,i_2} \hat p_{j_2}+ {\upsilon}_2] f^{-{1\over 2}}.
\end{eqnarray}

In order to satisfy Eq. (\ref{q12}) the commutator should not contain quadratic
terms in the momenta, so
$$  [-{i\over 2} f^{-1} (g^{i_1j_1} f)_{,i_1} \hat p_{j_1},\frac{1}{2}
g^{i_2j_2} \hat p_{i_2} \hat p_{j_2}]\ \ \text{and}\ \ [-{i\over 2} f^{-1}
(g^{i_2j_2} f)_{,i_2} \hat p_{j_2},\frac{1}{2}g^{i_1j_1} \hat p_{i_1} \hat
p_{j_1}] $$
should be zero. This can be achieved by demanding that $f(q^i)$ factorizes as
\begin{equation}\label{condfvol}
f(q^i)=f_1(q^{i_1})f_2(q^{i_2}).
\end{equation}
It should be remarked that the additional condition (\ref{condfvol}) is fully
compatible with (\ref{div6}).

As a consequence of the property (\ref{condfvol}), the term
$$[-{i\over 2} f^{-1} (g^{i_1j_1} f)_{,i_1} \hat p_{j_1},-{i\over 2} f^{-1}
(g^{i_2j_2} f)_{,i_2} \hat p_{j_2}]$$
is also zero.

Then it remains
\begin{equation}\label{g1g2c}
[\hat{\cal H}_1,\hat{\cal H}_2]=i c_{12}^af^{1\over 2} \hat{\cal H}_a f^{-
{1\over 2}}-f_1^{-1} (g^{i_1j_1} \upsilon_{2,j_1}f_1)_{,i_1}+ f_2^{-1}
(g^{i_2j_2} \upsilon_{1,j_2}f_2)_{,i_2},
\end{equation}
where we used the classical relation (\ref{chhcond}) in order to rebuilt
$\hat{\cal H}_a$. The last two terms in Eq. (\ref{g1g2c}) can be rewritten by
using the classical relations (\ref{vcg61}) and (\ref{vcg62})
 \begin{equation}\label{condvol}
  f_1^{-1} (c_{12}^a\xi_a^{i_1}f_1)_{,i_1}+f_2^{-1} (
c_{12}^a\xi_a^{i_2}f_2)_{,i_2}=c_{12}^a  f^{-1}
(\xi_a^{i}f)_{,i}=c_{12}^a(
C_{ab}^b+c^1_{a1}+c^2_{a2}),
\end{equation}
to cancel out as a consequence of the Jacobi identity (\ref{idj}).

To complete the proof, let us now evaluate the Eq. (\ref{q1a}):
\begin{eqnarray}\label{ccqlin1}
\lefteqn{\!\!\!\! [\hat{\cal H}_A,\hat{\cal H}_a]=[\frac{1}{2} f^{-{1\over 2}}
\hat p_{i_A} g^{i_Aj_A} f \hat p_{j_A} f^{-{1\over 2}}+ {\upsilon}_A,f^{1\over
2} \xi^k_a \hat p_k f^{-{1\over 2}}]} \nonumber \\ && \ \ \ \ \
=\frac{1}{2}f^{-{1\over 2}}( \hat p_{i_A} g^{i_Aj_A} f \hat p_{j_A}\xi^k_a \hat
p_k - \xi^k_a \hat p_k \hat p_{i_A} g^{i_Aj_A} f \hat p_{j_A} )f^{-{1\over
2}}+i \xi^k_a \upsilon_{A,k}\nonumber \\ && \ \ \ \ \ =\frac{1}{2}f^{-{1\over
2}}\left(i \hat
p_{i_A}(g^{i_Aj_A}_{,k_A}\xi_a^{k_A}-2g^{i_Ak_A}\xi_{a,k_A}^{j_A })\hat
p_{j_A}+ f g^{i_Aj_A} {[ \xi_a^k (ln f)_{,k}]_{,i_A}+ \xi_{a,i_Ak}^k }\hat
p_{j_A} \right)f^{-{1\over 2}}\nonumber \\ && \ \ \ \ \ \ \ +i \xi^k_a
\upsilon_{A,k}.
\end{eqnarray}
After using Eqs. (\ref{cond611}),(\ref{cond612}) or (\ref{cond621}),
(\ref{cond622}) one obtains
\begin{eqnarray}\label{ccqlin2}
\lefteqn{\!\!\!\!\!\!\!\!\!\!\!\! [\hat{\cal H}_A,\hat{\cal H}_a]=i c_{Aa}^A
\frac{1}{2}f^{-{1\over 2}}\hat p_{i_A}g^{i_Aj_A}f\hat p_{j_A}f^{-{1\over 2}} +i
\xi^k_a \upsilon_{A,k}} \nonumber \\ && \ \ +\frac{1}{2} f^{-{1\over 2}}\left(
f g^{i_Aj_A} {[ \xi_a^k (ln f)_{,k}]_{,i_A}+ \xi_{a,i_Ak}^k}\hat p_{j_A}
\right)f^{-{1\over 2}}.
\end{eqnarray}
The first two terms in (\ref{ccqlin2}) are $i c_{Aa}^{(A)} \hat{\cal H}_{(A)}$
(there no sum over label $A$). Then we have to see whether the last term is
zero. Actually,
\begin{equation}\label{anula}
[\xi_a^k (ln f)_{,k}]_{,i_A}+ \xi_{a,i_Ak}^k= [\xi_a^k (ln f)_{,k}+
\xi_{a,k}^k]_{,i_A}= [f^{-1}(\xi_a^k f)_{,k}]_{,i_A}=C_{a\beta,i_A}^{\beta}=0.
\end{equation}
Thus the demonstration is completed. $\hat\Omega$ is Hermitian
thanks to the choice (\ref{div6}) and is nilpotent due to the
factorization (\ref{condfvol}).

\section{Unitary transformation and constraint operators}

The former results can be generalized, whenever one takes into account
transformations leaving the BRST system invariant. In fact, the essential
properties of $\hat\Omega$ (Hermiticity and nilpotency) are not modified under
a unitary transformation

\begin{equation}
\hat \Omega  \rightarrow  e^{i\hat  C}~\hat\Omega~
e^{-i\hat C},\label{tu6}
\end{equation}
This transformation defines a new set of first class
constraint operators. By
choosing
\begin{eqnarray}\label{de6}
\lefteqn{\hat C={1\over 2}[\hat\eta^1 ~ F_1(q^i) ~ \hat{\cal P}_1-\hat{\cal
P}_1 ~ F_1(q^i) ~ \hat\eta^1 + \hat\eta^2 ~ F_2(q^j) ~ \hat{\cal P}_2-\hat{\cal
P}_2 ~ F_2(q^j) ~ \hat\eta^2] }\nonumber \\ &&=\hat\eta^1 ~ F_1(q^i) ~
\hat{\cal P}_1+  \hat\eta^2 ~ F_2(q^j) ~ \hat{\cal P}_2 +
\frac{i}{2}F_1(q^i)+\frac{i}{2} F_2(q^i) \nonumber \\ &&=-\hat{\cal P}_1 ~
F_1(q^i) ~ \hat\eta^1 -\hat{\cal P}_2 ~ F_2(q^j) ~ \hat\eta^2 -
\frac{i}{2}F_1(q^i) - \frac{i}{2}F_2(q^i) ,~~~~~~F_{1,2}(q)>0,
\end{eqnarray}
and using the identities
\begin{equation}
e^{i\hat\eta^A ~ F_A(q^i) ~ \hat{\cal P}_A} = 1+ i
\hat\eta^A ~(
e^{F_A(q^i)}-1) ~ \hat{\cal P}_A,
\end{equation}
\begin{equation}
e^{i \hat{\cal P}_A~ F_A(q^i) ~\hat\eta^A } = 1+ i
\hat{\cal P}_A ~(
e^{F_A(q^i)}-1) ~\hat\eta^A,
\end{equation}
one obtains
\begin{eqnarray} \lefteqn{\hat\Omega={\hat\eta}^1
e^{\frac{F_1-F_2}{2}}\hat {\cal H}_1
e^{\frac{F_1+F_2}{2}}+{\hat\eta}^1{\hat\eta}^2 ie^{\frac{F_1+F_2}{2}}[\hat
{\cal H}_1,e^{-F_2}] e^{\frac{F_1+F_2}{2}}\hat{\cal P}_2}\nonumber\\ &&~~+
{\hat\eta}^2 e^{\frac{F_2-F_1}{2}}\hat {\cal H}_2 e^{\frac{F_1+F_2}{2}}
+{\hat\eta}^2{\hat\eta}^1 ie^{\frac{F_1+F_2}{2}}[\hat {\cal H}_2,e^{-F_1}]
e^{\frac{F_1+F_2}{2}}\hat{\cal P}_1 \nonumber\\
&&~~+\frac{1}{2}{\hat\eta}^1{\hat\eta}^2 e^{F_1+F_2}\hat c_{12}^a \hat{\cal
P}_a + {\hat\eta}^a e^{-\frac{F_1+F_2}{2}}\hat {\cal H}_a e^{\frac{F_1+F_2}{2}}
\nonumber \\&&~~+ {\hat\eta}^1{\hat\eta}^a \left(\hat
c_{1a}^1-ie^{\frac{F_1-F_2}{2}}[\hat {\cal H}_a,e^{-F_1}]
e^{\frac{F_1+F_2}{2}}\right)\hat{\cal P}_1 \nonumber\\&&~~
+{\hat\eta}^2{\hat\eta}^a \left(\hat c_{2a}^2-ie^{\frac{-F_1+F_2}{2}}[\hat
{\cal H}_a,e^{-F_2}] e^{\frac{F_1+F_2}{2}}\right)\hat{\cal P}_2+\frac{1}{2}
{\hat\eta}^a{\hat\eta}^b C_{ab}^c\hat{\cal P}_c \label{omegaescf}\end{eqnarray}


Thus one can identify constraint operators and structure
functions in Eq.
(\ref{omegaescf}),
\begin{equation}\hat H_1={1\over 2}
e^{\frac{F_1-F_2}{2}}f_1^{-{1\over 2}}\hat p_{i_1}g^{i_1j_1}f_1 \hat p_{j_1}
f_1^{-{1\over 2}}e^{\frac{F_1+F_2}{2}}+e^{F_1}{\upsilon}_1,
\label{vcuadq'v1}\end{equation}

\begin{equation}
\hat H_2={1\over 2} e^{\frac{F_2-F_1}{2}}f_2^{-{1\over 2}}\hat p_{i_2}
g^{i_2j_2}f_2 \hat p_{j_2} f_2^{-{1\over
2}}e^{\frac{F_1+F_2}{2}}+e^{F_2}{\upsilon}_2,\label{vcuadq'v2}\end{equation}

\begin{equation}\hat {\cal H}_a=
e^{-\frac{F_1+F_2}{2}}f^{1\over 2} \xi^i_a\hat p_i f^{-{1\over
2}}e^{\frac{F_1+F_2}{2}}. \label{vlinq6'}\end{equation} The resulting
constraint operators, Eqs. (\ref{vcuadq'v1}),(\ref{vcuadq'v2}), correspond to
scaled super-Hamiltonian constraints
 $H_A=e^{F_A} {\cal H}_A$. One can name  $G^{i_Aj_A}=
e^{F_A} g^{i_Aj_A},~ V_A= e^{F_A}{\upsilon}_A$ to write

\begin{equation}\hat H_1={1\over 2}
e^{\frac{F_1-F_2}{2}}f_1^{-{1\over 2}}\hat
p_{i_1}
e^{-F_1} G^{i_1j_1}f_1 \hat p_{j_1} f_1^{-{1\over
2}}e^{\frac{F_1+F_2}{2}}+V_1,
\label{vcuadq'v61}\end{equation}

\begin{equation}\hat H_2={1\over 2}
e^{\frac{F_2-F_1}{2}}f_2^{-{1\over 2}}\hat p_{i_2} e^{-F_2} G^{i_2j_2}f_2 \hat
p_{j_2} f_2^{-{1\over 2}}e^{\frac{F_1+F_2}{2}}+V_2,
\label{vcuadq'v62}\end{equation} with the corresponding structure functions
\begin{eqnarray}\label{newssf}
\lefteqn{\hat  C_{AB}^B= i e^{\frac{F_1+F_2}{2}}[\hat{\cal
H}_A,e^{-F_B}]
e^{\frac{F_1+F_2}{2}}} \nonumber\\
&&~~=\frac{1}{2}e^{\frac{F_A}{2}}
f^{-{1\over2}}[\hat p_{i_A} f g^{i_Aj_A} F_{B,j_A} +
F_{B,j_A} f
g^{i_Aj_A}\hat p_{j_A}] f^{-{1\over 2}} e^{\frac{F_A}{2}},
\end{eqnarray}
\begin{equation}\hat C_{12}^a= e^{\frac{F_1+F_2}{2}}\hat
c_{12}^a,\end{equation}
\begin{equation}\hat C_{Aa}^A=\hat c_{Aa}^A+\xi^i_a
(F_A)_{,i},\end{equation}
\begin{equation}\hat C_{ab}^c=\hat C_{ab}^c.\end{equation}
All the operators and structure functions are ordered in such a way they
satisfy,
\begin{equation}[\hat  H_1,\hat  H_2]=\hat C_{12}^1(q,p)
\hat H_1  +  \hat
C_{12}^2(q,p) \hat H_2+\hat C_{12}^a\hat{\cal H}_a,
\label{ccqf}\end{equation}
\begin{equation}[\hat  H_A,\hat  {\cal H}_a]=\hat
C_{Aa}^{(A)} \hat H_{(A)},
\label{ccqfnw}\end{equation}
\begin{equation}[\hat {\cal H}_a,\hat {\cal H}_b]=\hat
C_{ab}^c \hat {\cal H}_c \label{clqf}\end{equation} (there is no sum over label
$A$), i.e., the algebra is free from anomalies at the quantum level.

Of course, the scaling of constraints does not modify the dynamics of the
system. However the constraint algebra looks different: the Eq. (\ref{ccqfnw})
now involves all constraints and the structure functions are no longer
constant. Since the only consequence of the proposed unitary transformation is
the scaling of the super-Hamiltonians, the commutators among supermomenta
remain unchanged.

\section{Physical inner product}

To complete the quantization it is necessary to define a physical inner product
where the spurious degrees of freedom are frozen by means of gauge fixing
conditions (see also Ref. \cite{bl}). Here it is relevant to take into account
the role played by the invariance transformations of the theory: (i) general
coordinate transformation, (ii) linear combinations of supermomenta
constraints, (iii) scaling of super-Hamiltonian constraints. The physical inner
product for Dirac wave functions
\begin{equation}
(\varphi_1,\varphi_2)=\int dq\ \big[\prod^{m+2} \delta(\chi)\big]\ J\
\varphi^*_1(q)\ \varphi_2(q)\label{prodesc6}\end{equation} (where $J$ is the
Faddeev-Popov determinant and $\chi$ represents $m+2$ gauge conditions) must be
invariant under any of these transformations. By regarding the behavior of the
constraint operators under transformations (i-iii) one realizes that the Dirac
wave functions should transform according to \cite{fs97}
\begin{equation}\label{unmedio6}
\varphi  \rightarrow    \varphi'=(det    A)^{1\over    2}
e^{-{F_1+F_2\over 2}}
\varphi,
\end{equation}
$A$ being the matrix of the combination of linear
constraints. Therefore the
Faddeev-Popov determinant $J$ in the physical inner product
should change in opposite way, in order that the inner
product remains
unchanged.

In Eq. (\ref{prodesc6}) there are $m$ functions $\chi$ fixing $m$
coordinates associated to the supermomenta constraints, whose
characteristics are very well known. So, let us pay attention to
the remaining two gauge fixing functions, which come from the
super-Hamiltonians and are involved with the scaling
transformation. As it was shown in previous works, the scaling
factor can be associated either with a positive definite potential
(intrinsic time case \cite{fs97}) or with the norm of a conformal
Killing vector (extrinsic time case \cite{fs99}). Up to this point
it has not been necessary to make any assumption on the potentials
of the super-Hamiltonians (except for their functional
dependence). However, some care about the type of potential must
be taken to fix the gauge. If the potentials are definite
positive, the time is intrinsic and the scaling factor of the
super-Hamiltonians will be the potentials themselves: $F_A= ln
V_A$ and the associated gauge condition will be the one studied in
Ref. \cite{fs97}. If the time is extrinsic, the scale factors are
associated with the norm of a conformal Killing vector for each
super-Hamiltonian:  $F_A= ln |\vec\xi_A|^{-2}$ and the inner
product will be the one defined in Ref. \cite{fs99}.

However, there are $m+2$ constraints in the theory, so we need one more gauge
condition. Since, this finite dimensional model mimics the constraint algebra
of general relativity, the two subspaces $(q^{i_1})$ and $(q^{i_2})$ are
interpreted as the field at two different points of the space-time, and so the
super-Hamiltonians ${\cal H}_1$ and ${\cal H}_2$ are regarded as ``the
super-Hamitonian'' evaluated in two diferent points in space-time. Then, we
suppose that both super-Hamiltonian have the same type of time, and in fact,
both contain the same time. So, we propose as the remaining gauge condition
\begin{equation}\label{condgaugtemp}
\chi=\delta(t_1-t_2).
\end{equation}
This gauge condition can be naturally retrieved in the framework of the
multiple time formalism \cite{lo86,lo89,lu97}. Therefore, in both cases
-intrinsic and extrinsic time- the inner product is regularized by including
Eq. (\ref{condgaugtemp}) in the already known set of $m+1$ gauge conditions.

\section{Conclusions}

In this work we managed to naturally extend the previous ordering findings of
Refs. \cite{fs97,fs99,fs00} to the interesting case of systems subject to more
than one super-Hamiltonian constraint. As long as we know, this is the first
work that treat ordering problems in such case and it also prove the power of
the BRST formalism for providing the necessary tools for the treatment of
invariance properties at the quantum level. By taking advantage of these
invariance transformations of the theory, in particular the scaling invariance
of the super-Hamiltonians, we were able to raise a nontrivial algebra between
the super-Hamiltonians and to find the anomaly-free ordering, namely Eqs.
(\ref{ccqf})-(\ref{clqf}). In this case, each scaling independently contribute
with a term, involving also the momenta constraints. The role played by the
invariance transformations, and how they modify the operator ordering, has been
briefly discussed in Sec. V (see also Refs. \cite{fs97,fs99} for a more in
depth analysis).

When a finite dimensional system is proposed as a model for quantizing a
general covariant theory (such as for example, general relativity), always
remains the suspicion about the real competence of the model when it is
compared with the infinite dimensional case. However, we can learn something
about the ordering of the constraint operators that must be applied to both
cases: the invariance transformation of the theory substantially modify the
ordering. For example, if one admits that the super-Hamiltonian in the infinite
dimensional case can be scaled, then the ordering for the supermomenta cannot
be a simple functional derivation with respect to the canonical field variable,
as usually appears in the literature, but it must ``wear" the scaling factors
in both sides of it.

\acknowledgments

This research was supported by Universidad de Buenos Aires (Project No. TX 64)
and Consejo Nacional de Investigaciones Cient\'\i ficas y T\'ecnicas. D. M. S.
thanks the members of the Department of Physics and Astronomy of the University
of Pittsburgh, specially Roberto Gomez, for kindly providing facilities which
made possible to finish this work.

\end{document}